# Toward the Autonomous AI Doctor: Quantitative Benchmarking of an Autonomous Agentic AI Versus Board-Certified Clinicians in a Real World Setting

Hashim Hayat,[1] Maksim Kudrautsau,[1] Evgeniy Makarov,[1] Vlad Melnichenko,[1] Tim Tsykunou,[1] Piotr Varaksin,[1] Matt Pavelle,[1] and Adam Z. Oskowitz, MD, PhD[1,2]

[1]Doctronic Research Group, New York, NY

[2]Department of Surgery, University of California San Francisco, San Francisco, CA

**Background:** Globally we face a projected shortage of 11 million healthcare practitioners by 2030, and administrative burden consumes 50% of clinical time. Artificial intelligence (AI) has the potential to help alleviate these problems. However, no end-to-end autonomous large language model (LLM)-based AI system has been rigorously evaluated in real-world clinical practice. In this study, we evaluated whether a multi-agent LLM-based AI framework can function autonomously as an AI doctor in a virtual urgent care setting.

**Methods:** We retrospectively compared the performance of the multi-agent AI system Doctronic and board-certified clinicians across 500 consecutive urgent-care telehealth encounters. The primary end points: diagnostic concordance, treatment plan consistency, and safety metrics, were assessed by blinded LLM-based adjudication and expert human review.

**Results:** The top diagnosis of Doctronic and clinician matched in 81% of cases, and the treatment plan aligned in 99.2% of cases. No clinical hallucinations occurred (e.g., diagnosis or treatment not supported by clinical findings). In an expert review of discordant cases, AI performance was superior in 36.1%, and human performance was superior in 9.3%; the diagnoses were equivalent in the remaining cases.

**Conclusions:** In this first large-scale validation of an autonomous AI doctor, we demonstrated strong diagnostic and treatment plan concordance with human clinicians, with AI performance matching and in some cases exceeding that of practicing clinicians. These findings indicate that multi-agent AI systems achieve comparable clinical decision-making to human providers and offer a potential solution to healthcare workforce shortages.

**Visual Abstract**

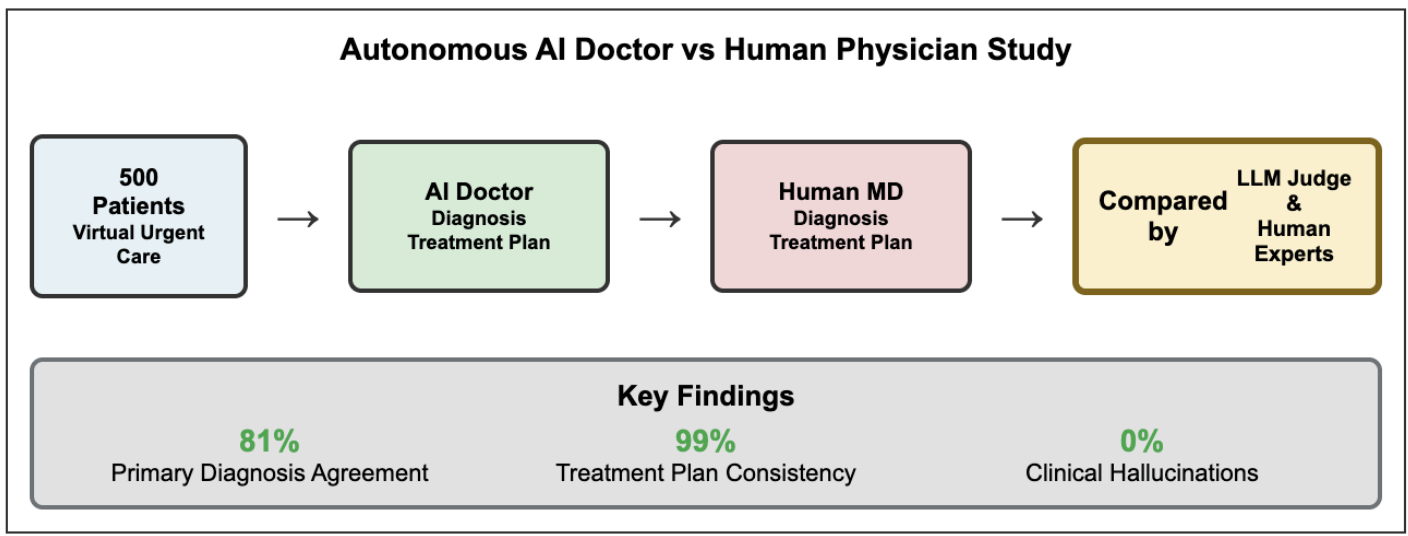

## Introduction

The global healthcare workforce faces a critical shortage of providers. By some estimates, an additional 4.3 million practitioners are currently needed to meet the demand for basic health service worldwide, and this deficit is predicted to reach 11 million by 2030 [1]. The World Health Organization estimates that 8 million lives are lost annually due to treatable disease [2]. The United States alone projects a shortage of 124,000 physicians by 2034, including 48,000 in primary care and 77,000 in specialty medicine [3]. This crisis will only intensify as demographic transitions drive exponential growth in demand: populations aged 65+ will triple globally by 2050 [4]. Simultaneously, the burden of care and documentation has increased. Administrative burden consumes 50% of clinical time, driving burnout rates above 45.8% across specialties [5]. Traditional solutions: medical school expansion, telemedicine deployment, and workflow optimization cannot bridge this gap within viable time frames, as physician training requires 11-15 years and faces enrollment constraints.

Large language models (LLMs), including foundational models, and autonomous AI systems (the first technological solution capable of exceeding human capacity limitations in clinical decision-making and management) could potentially address both supply constraints and rising demand [6,7]. With advances in natural language processing and generative reasoning, LLMs have rapidly evolved from documentation assistants to sophisticated agents capable of multi-step clinical reasoning, differential diagnosis, and evidence-based management [8-10].

In a blinded, expert-rated vignette trial, the accuracy of differential diagnosis and management planning by the autonomous medical LLM agent Articulate Medical Intelligence Explorer (AIME) was similar to that of practicing physicians [11]. This study established methodological standards for real-world representativeness, blinding, and rigorous error coding. Our investigative approach builds on this foundation. But rather than relying on thousands of heterogeneous, authentic patient records as benchmarks, we focused on transparent, reproducible prompt-based adjudication of real-world interactions. More recently, a proprietary, non-LLM-based AI intake and recommendation system developed by K Health was evaluated in a real-world virtual urgent-care setting at Cedars-Sinai. The system engaged patients in structured chat before their video visits, producing diagnostic and management suggestions-including prescriptions, laboratory tests, and referrals-based on symptom input and EHR data. In a retrospective study of 461 adult patient visits with respiratory, urinary, vaginal, eye, or dental symptoms, AI and clinician recommendations were concordant in 56.8% of cases. Expert adjudicators rated the AI's recommendations as optimal more often than those of clinicians (77.1% vs. 67.1%), and AI outperformed clinicians in guideline adherence and recognition of red flags [12].

Nevertheless, critical gaps persist in the evaluation and use of fully autonomous clinical AI systems. Notably, an end-to-end LLM-based agentic AI system has never been evaluated clinically. In addition, most current systems do not provide end-to-end autonomy; they assist or suggest, rather than independently reason, decide, and document the full encounter. Recent studies also do not provide unbiased clinically relevant rigorous evaluation; published benchmarks often rely on simulated, vignette, or narrow specialty cohorts of <500 cases. They do not rely on real-world, demographically diverse patient flows. Another clinical shortcoming is the lack of reproducible error coding, as few studies apply granular, consensus-driven error taxonomies, such as omissions, hallucinations (e.g., diagnosis/treatment not supported by clinical findings), and inappropriate plans in blinded adjudication.

Essential questions remain. Can a multi-agent framework that uses LLMs and foundational models function as an AI doctor? Specifically, can it perform all the tasks needed to function autonomously, including history taking, conversational synthesis of patient information, clinical reasoning, guideline-concordant management, and detailed documentation, all without direct human-in-the-loop intervention? Can a fully autonomous AI doctor safely perform these complex and nuanced clinical tasks in real-world settings? What are the strengths, weaknesses, and risks of AI-driven versus human clinician decision-making in real-world settings? How do we ensure robust benchmarking, operational safety, transparency, and reproducibility as such systems move from the lab to patient encounters?

To answer these questions, we rigorously compared Doctronic, a proprietary, multi-agent, LLM-based clinical reasoning and documentation system, against experienced human clinicians in real-world virtual urgent care encounters. Using blinded expert and LLM-based adjudication, explicit error taxonomies, and robust reproducibility frameworks, we assessed (1) diagnostic concordance and safety; (2) adherence to management guidelines; (3) depth, clarity, and consistency of medical documentation; (4) frequency, type, and risk level of clinical errors with a focus on hallucinations; and (5) the generalizability and operational implications for healthcare workforce and equity. The goals of the study were to define the technical capabilities of an agentic AI in real-world medical practice and to establish new standards for transparent, reproducible clinical AI benchmarking. We tested three hypotheses: (1) that AI diagnostic/treatment decisions are consistent with those of board-certified clinicians when evaluating the same patients and clinical scenarios; (2) that AI-driven documentation can improve consistency and efficiency, with minimal factual hallucinations; and (3) that major clinical errors by AI agents can be reduced to near zero with the appropriate architecture and grounding.

## Materials and Methods

*Study Design and Setting*

We conducted a retrospective, observational study, analyzing 500 consecutive, fully de-identified (no linkage for re-identification), urgent care telehealth encounters occurring in the first week of March 2025. Each encounter was conducted in sequence: first by Doctronic, our proprietary multi-agent AI doctor, operating autonomously, and then by a board-certified clinician. Clinicians were permitted, but not required, to use the AI-based documentation during their evaluation.

*Doctronic*

Doctronic is a cloud-native modular system that has over 100 LLM-powered agents, each with a distinct, well-defined clinical role, mirroring the structured responsibilities of a human care team. These agents operate cooperatively, passing context-rich data to one another dynamically. The entire system is designed to mimic the clinical tasks of a primary care doctors office. Within this context, Doctronic is capable of performing a full medical history, after which it generates a SOAP note with the following components 1) a summary of the HPI including self reported physical findings and diagnostic tests, 2) a differential diagnosis with at least 4 diagnosis, 3) A plan for further diagnostic evaluation and treatment based on the differential. For the patient, the experience is similar to a two-way, open-ended, text-based chat.

*Patient Population and Case Selection*

Cases were drawn from routine, nonemergent telehealth encounters in the U.S. The study cohort was demographically and clinically heterogeneous, reflecting the prevailing mix of modern urgent-care cases, including acute respiratory, dermatological, musculoskeletal, and gastrointestinal complaints and infections. De-identified patient transcripts, including AI chat logs and AI- and clinician-generated documentation and treatment plans, served as input. Exclusion criteria included incomplete transcripts, duplicate encounters, or lack of both AI and clinician notes for the same encounter.

*Clinicians*

The performance of Doctronic was compared with that of board-certified, U.S.-licensed clinicians that specialize in telehealth. This included physicians and appropriately trained Advanced Practice Providers. The clinicians were trained on the Doctronic system prior to the study. Clinicians were given a copy of the AI-generated documentation before their telehealth visit with the patient. The video-based clinical encounters followed standard telemedicine protocols, enabling the clinician to gather as much or as little additional information as needed. SOAP notes (subjective, objective, assessment, and plan) were used for blinded adjudication (see Appendix 1 for examples).

*Evaluation and Benchmarking*

To enable transparent, reproducible, and meaningful comparison of the performance of Doctronic vs. human clinicians, we analyzed three core metrics: (1) the consistency of the diagnoses, determined by blinded LLM-based judge prompts; (2) the consistency of treatments and plans, assessed in both binary fashion (compatible/incompatible) and on an ordinal scale (0-10); and (3) Surface-level textual similarity followed by contextual and semantic similarity of the documentation, along with the presence or absence of clinical hallucination. For the initial adjudication, we used a consensus multi-run LLM-judge protocol with pre-tested prompts.

*Blinded LLM-Judge Prompts*

Prompts were developed and validated on a set of paired notes generated by Doctronic and the human clinicians. Randomly selected paired notes were evaluated by human judges to ensure fidelity to the prompt's purpose [13]. We constructed and validated four primary LLM prompts for automated, reproducible adjudication (Appendix 2). (1) Top-4 concordance: Does at least one of the top four diagnoses of the LLM match or correspond to that of the clinician (with semantic allowance for clinical synonyms)? (2) Top-1 concordance: Does the primary diagnosis of the LLM match that of the clinician (with semantic allowance for clinical synonyms). (3) The safety and compatibility of the treatment plan: Are the treatment plans and recommendations consistent and safe, accounting for potential synonyms, equivalent formulations, or minor differences? (4) Comparative summary score (CSS): Were the diagnostic thoroughness, rationale, and decision quality of the LLM and clinician similar as rated on a 0-10 scale? The CSS was accompanied by a natural language explanation of the scores. The LLM judge role used GPT-4.0 by OpenAI.

*Evaluation by Human Experts*

Each encounter pair in which the top diagnosis of AI and clinician did not match was evaluated by a board-certified physician with access to medical reference material. Blinding the physician to the origin of the documentation proved impractical, as the AI-based notes were highly consistent and thus easily recognized within a few pairs. The physician was asked to determine the cause of the disagreement between the documents, whether AI or the physician was more likely to be correct, whether it was not possible to determine which diagnosis was more appropriate, and whether the diagnoses did, in fact, match.

*Similarity and Style Metrics*

To evaluate how similar-or different the AI-generated (Doctronic) and clinician-generated SOAP notes were, we followed a two-step process. First, we assessed surface-level textual similarity using three standard statistical metrics: (1) TF-IDF cosine similarity, which transforms each note into a weighted term-frequency vector and measures the cosine of the angle between them to capture word-frequency alignment; (2) the Jaccard index, which is the ratio of the intersection to the union of lowercased token sets, ranging from 0 (no overlap) to 1 (identical token sets); and (3) the Levenshtein ratio, a normalized edit-distance score based on character-level insertions, deletions, and substitutions that quantifies textual similarity on a 0-1 scale. These analyses demonstrated only minimal alignment in phrasing, formatting, and vocabulary. Then, to probe contextual and semantic similarity, we generated embeddings for each note using OpenAI's text-embedding-3-small model and two versions of Biobert, a domain-specific adaptation of BERT pre-trained on biomedical corpora. For these, we computed cosine similarity between embeddings to evaluate whether the notes conveyed equivalent clinical meaning, regardless of surface differences. All computations were implemented with a custom evaluation script in Python 3.12 and applied independently to each Doctronic-clinician pair, offering a reproducible framework that distinguishes stylistic divergence from semantic alignment.

*Statistical Analysis*

All statistical analyses were done in Python (using NumPy, Pandas, Scikit-learn, and SciPy) and R. Values are reported as mean ± SD. We summarized categorical endpoints including AI-physician concordance, discordance rates, and occurrence of hallucinations as counts and percentages. No formal hypothesis tests or a priori power analysis were performed, as this was an exploratory evaluation.

**RESULTS**

**Table 1. Summary Demographics and Presenting Complaints**

| Characteristic | Value |
| --- | --- |
| Demographics | |
|   No. of encounters | 500 |
|   Median age, years (range) | 39 (18-88) |
|   Female/male | 57%/42% |
| Most common clinical conditions | |
|   Acute sinusitis | 25 |
|   Influenza | 20 |
|   Acute bronchitis | 17 |
| Co-morbidity of 2+ chronic conditions | 53% |

The analytic dataset consisted of 500 encounter pairs (one AI, one human per encounter). The types of cases, patient demographics, and presenting complaints (Table 1) reflected broad, real-world distribution of urgent care cases. The sample was balanced for age, gender, and case complexity based on co-morbidity, and the presenting problems spanned over 100 major diagnostic categories in ICD-10 (Appendix 3). The median session length was 8.4 min (interquartile range [IQR] 5.1-11.7) for AI.

*Clinical Agreement and Safety*

The primary safety end point was the consistency of the treatment plan, specifically if treatment plans represent compatible clinical management strategies that would lead to similar therapeutic outcomes. In 496 of the 500 encounter pairs (99.2%; 95% confidence interval [CI] 98.1-99.8%), the management plans were judged to be clinically compatible and guideline-concordant by blinded LLM judge. In four encounter pairs (0.8%) the LLM showed substantive divergence (e.g. the AI system recommending a significantly broader diagnostic work-up for complex symptoms), but none entailed immediate or high-harm risk.

*Diagnostic Concordance, Error/Hallucination Frequency, CSS*

The top primary diagnosis of AI and the clinician matched In 81% of cases (405/500; 95%CI: 77.1-84.5%). At least one of the four most likely diagnoses matched in 95.4% of cases (477/500; 95%CI: 93.3-97.0%). Technical errors occurred in one case where AI-generated notes lacked sufficient patient data needed for evaluation. However, there were no cases of hallucination, in which AI notes contained a clinically fabricated diagnosis or a treatment plan not supported by the transcript (one-sided 95%CI: up to 0.7% allowed).

**Table 2. Comparative Summary Scores by Encounter Complexity**

| Complexity | CSS (mean) | No. of cases |
| --- | --- | --- |
| Low (<3) | 6.1 | 11% |
| Moderate (4,5,6) | 6.1 | 86% |
| High (>6) | 6.4 | 3% |

To evaluate the alignment between AI-generated and clinician-generated SOAP notes, we also used a structured metric called the clinical similarity score (CSS). The CSS is a four-part composite LLM-as-a-judge that intends to identify (1) the degree of clinical agreement between the two notes, (2) the complexity of the patient's condition, (3) the presence of co-morbidities, and (4) the primary diagnosis based on International Classification of Diseases (ICD) terminology. The first component, scored 0-10, measures how similarly a clinician would interpret the case based on each note, with 10 indicating nearly identical clinical impressions and 0 a complete mismatch. The second component, also measured on a 0-10 scale, reflects the inherent medical complexity of the case-with common, self-limiting conditions like influenza scoring low, and serious or multisystem diseases like polymyositis scoring high. The co-morbidity flag indicates whether multiple clinically relevant conditions are present, and the ICD label standardizes the main clinical diagnosis for comparison and downstream coding. The CSS stratified by case complexity is shown in Table 2. Cohort-wide, the average CSS assigned to AI-vs-human notes by the LLM judge was 6.1 ± 1.5 (range 1-9).

For instance, in a high-similarity case (CSS: 8/10; chief complaint: bacterial vaginosis), both the AI- and clinician-generated SOAP note agreed closely on the diagnosis and management, with only minor differences in the breadth of differential diagnosis. A medium-similarity example (CSS: 6/10; chief complaint: genital herpes simplex) showed agreement on the primary diagnosis, but Doctronic's note was more comprehensive, incorporating secondary conditions like postherpetic neuralgia, which the clinicians note did not address. In contrast, a low-similarity example (CSS: 3/10; chief complaint: urinary tract infection, site not specified), AI offered a broad differential diagnosis not linked to key symptoms, whereas the clinician provided a single diagnosis and a more definitive treatment plan.

*Human Review of Discordant Pairs*

To further understand the clinical significance of discordances between Doctronic and human clinicians, a board-certified physician manually reviewed all 97 encounter pairs in which the blinded LLM judge did not find primary diagnostic concordance (Top-1 match). Each case was assessed for the completeness, logic, and appropriateness of clinical reasoning and overall clinical safety.

In 35 cases (36.1%), the AI-generated note was judged to demonstrate more appropriate clinical reasoning or adherence to guideline-based care than the clinician-generated note, often in scenarios with established diagnostic or management protocols, where the AI's consistency and attention to up-to-date guidelines provided a clear advantage. Also included in this group were cases in which the diagnoses were atypical at an urgent care center, such as low-acuity neurological symptoms or complaints spanning multiple systems.

In 9 cases (9.3%), the clinician's note was rated superior, typically in scenarios where the patients presented with known patterns of disease but also had additional distracting complaints. In these cases, the clinician appeared more comfortable disregarding these symptoms and more frequently documented the reasoning for differential diagnoses.

In 36 cases (37.1%), the primary diagnosis by AI and the clinician were the same, but not recognized as such by the LLM judge, owing to documentation by the clinician with low specificity. In one example, the primary diagnosis was listed as a viral infection by the physician and as viral pharyngitis by AI, although both had treatment plans focused on a viral upper respiratory tract infection. This finding of LLM not judging diagnosis as the same due to low specificity by the human SOAP note also contributed to the 19 cases where the top 4 diagnosis were not concordant, but the treatment plans did show alignment. In several examples the human SOAP note provided a presenting symptoms as the diagnosis, but did provide an appropriate treatment plan.

In 17 cases (17.5%), the available documentation was insufficient or ambiguous, and the expert could not confidently determine whether the clinician or AI was superior. These cases typically involved scenarios in which an argument could be made that the diagnosis was unclear or that both diagnoses were the top choice (e.g., contact dermatitis vs. mild cellulitis).

Thus, Doctronic's performance was equal or superior to that of the clinician in more than one-third of the discordant cases. Importantly, across all discordant pairs, AI-generated notes never introduced a potentially harmful error, underscoring the robust safeguards and alignment with best safety practices.

*Textual and Semantic Similarity*

**Table 3. Surface Similarity and Embedding-Based Cosine Semantic Similarity Metrics of AI- and Clinician-Generated SOAP Notes Pairs (n=500)**

| Metric | Mean ± SD |
| --- | --- |
| Surface similarity | |
|   TF-IDF | 0.2285 ± 0.084 |
|   Jaccard index | 0.087 ± 0.0450. |
|   Levenshtein ratio | 0.364 ± 0.054 |
| Embedding-based cosine semantic similarity | |
|   OpenAI | 0.693 ± 0.064 |
|   Biobert Base Cased-v1.1 | 0.891 ± 0.026 |
|   Biobert Base Cased-v1.2 | 0.889 ± 0.023 |

To quantify the similarity between AI-generated and clinician-authored SOAP notes, we applied three complementary text-based metrics, each reflecting different aspects of surface similarity (Table 3). TF-IDF cosine similarity assesses the importance of the words used. Term frequency (TF) measures how often each word appears in a note. Inverse document frequency (IDF) reduces the influence of words that are common across the dataset, emphasizing those unique to individual notes, highlighting the extent to which the two notes share the same patterns of important, distinguishing words. The mean TF-IDF score was low (mean 0.2285 ± 0.084), indicating that AI and human notes generally differed in key language features.

The Jaccard index, which assesses the proportion of shared unique words between notes, was also low (mean 0.087 ± 0.045), indicating that, on average, fewer than 9% of the words in a typical note pair were shared. Thus, the overlap in vocabulary between AI and human authors was minimal.

The Levenshtein ratio, which measures character-level similarity based on the minimum edits needed to transform one note into another, was also fairly low (0.364 ± 0.054), indicating a slightly higher similarity than the word-level metrics, but nevertheless reflected considerable differences in phrasing and formatting. In other words, despite documenting the same clinical cases, the notes generated by AI and clinician typically diverged in the specifics of wording and structure.

In contrast, the embedding-based cosine similarity, which captures higher-level semantic relationships, yielded substantially higher scores (Table 3). Thus, despite stylistic and lexical variations, the notes showed high semantic similarity as measured by embeddings.

These results indicate that although Doctronic and human authored notes often differed in exact wording and structure, they were semantically consistent in their documentation of clinical reasoning, assessment, and treatment plans. This pattern supports the potential for AI-generated documentation to accurately reflect the intent and substance of human-written medical notes, even when surface expressions diverge.

**Discussion**

In this study, the most extensive real-world validation of an autonomous, multi-agent AI doctor, we compared the performance of Doctronic with that of board-certified clinicians across 500 consecutive urgent-care telehealth encounters. Our findings confirmed all three of our hypotheses: (1) that diagnostic and treatment decisions of AI are consistent with those of board-certified clinicians when evaluating the same patients and clinical scenarios; (2) that AI-driven documentation can improve consistency and efficiency, with minimal factual hallucinations; and (3) that major clinical errors by AI agents can be reduced to near zero with the appropriate architecture and grounding. Specifically, we found that the top diagnosis of Doctronic and the clinician matched in more than 80% of cases, and the treatment plans aligned in more than 99% of cases. The robustness of the results across patients, complaints, and demographic subgroups was confirmed by qualitative and statistical analyses. In an expert review of 97 discordant cases by board-certified clinicians, the diagnostic performance of AI was judged to be superior in 36.1% of the cases, and human performance was judged to be superior in 9.3%. In the remaining cases, the diagnosis was either ambiguous or, in fact, the same but worded in substantially different language. Thus, Doctronic was as good as experienced clinicians in this setting and performed at or above the standard of published AI systems.

Unlike prior assistive systems [14] that only provided recommendations for clinicians to review, Doctronic independently conducted the full clinical evaluation, including history taking, reasoning, and documentation, without human-in-the-loop input. The principal clinical finding, that Doctronic can autonomously generate an accurate diagnosis and a safe treatment plan in virtually all cases, has direct implications for urgent care access and quality. Although AI notes occasionally omitted contextual detail or failed to include relevant differential diagnoses, these omissions almost never translated into unsafe plans or missed management actions. The absence of hallucinated facts (e.g., diagnosis/treatment not supported by clinical findings) further distinguishes Doctronic from older documentation LLMs and even some current hybrid AI scribes [14].

Our analysis of discordant cases yielded critical insights into how best to evaluate AI systems in a clinically relevant context. Medical diagnoses are frequently uncertain, and the role of the clinician, or in this case AI, is to form a comprehensive differential diagnosis and test the hypothesis that one of those diagnoses is correct, through diagnostic tests or an empirical trial of treatment. In either case, the critical function is to develop an appropriate evaluation and treatment plan. Like human clinicians, AI and a clinician may disagree about a diagnosis. Such cases should be evaluated on the bias of treatment plan consistency, a metric designed to determine whether the initial management plan will lead to safe, similar, and appropriate management in the context of uncertainty. By this metric, Doctronic and expert clinicians were nearly perfectly aligned.

Our findings suggest that, with structured safeguards, agentic AI systems can be used in sequence with a clinician or as a first-line tool to increase access to urgent care that would otherwise be delayed because of workforce shortages and wait times for medical care. For resource-constrained environments and after-hours medicine, autonomy of this quality can dramatically increase access and decrease wait times. In developed-world health systems, the main utility of AI is augmenting throughput and enable clinicians to focus on high- complexity, high-touch, or longitudinal relationships.

*Limitations*

Despite the strengths of this large-scale, real-world, blinded evaluation of an AI doctor, several limitations must be acknowledged. Perhaps most importantly, our study aimed to assess the safety of a multi-agent autonomous system in real-world settings using concordance not correctness. We chose this approach because what is "right" is sometimes ambiguous. Ground truth based on follow-up outcomes was not considered. Therefore, the findings show agreement, not accuracy. Another limitation is that the AI notes were given to the clinicians before each clinical encounter, potentially creating an anchoring effect that may have biased them to align with the AI-based evaluation. Moreover, the clinicians were able to gather additional information on top of that provided by AI. As with the so-called "attending effect", whereby senior doctors receive significantly different information than trainees from the same patient, the human clinicians may have received different information that changed their diagnosis and approach.

Other important limitations are related to the current investigation of multi-agent AI. First, our results reflect the U.S. and English-language based urgent care and may not be generalizable to non-English speakers and non-telehealth settings. Second, since we used an LLM as a judge, with human level inspections as a safeguard against systematic bias, the LLM judge might have introduced some level of error. LLM as a judge is an emerging methodology and has not been widely used until recently. On-going use of diverse models and human adjudicators is recommended.

In the future, the performance of clinicians without access to the AI note should be evaluated to mitigate the anchoring bias introduced when clinicians review AI-generated notes first. More complex evaluation systems to ensure multiple metric based evaluations will be critical in the next phase of testing. This should include post visit patient follow up to enable the use of patient specific outcomes to evaluate the effectiveness of treatment plans.

## Conclusion

A state-of-the-art multi-agent AI doctor (Doctronic) can autonomously and safely assess and provide an appropriately documented treatment plan for patients in an urgent care setting with a consistency matching that of board-certified clinicians. With proper safeguards, explicit auditability, and rigorous benchmarking, such systems hold promise as a viable strategy to address the global shortage of medical providers, expand healthcare access and quality, and support sustainable and equitable future models of care.

## Acknowledgments

We thank the clinical annotation team and the Doctronic software developers, prompt writers, and clinicians for their expertise, feedback, and access to de-identified note data. We are grateful to external reviewers for critical questioning of our error taxonomy. No proprietary or patient-sensitive information was used in this manuscript.

## Disclosures

All authors are equity owners of Doctronic.

# Appendix 1: Examples of SOAP note pairs

Case 1: This SOAP note pair received the following scores: Top diagnosis: concordant; top 4 diagnoses: concordant; treatment plan alignment: not concordant, CSS 3/10

| Doctronic SOAP Note | Human Clinician SOAP note |
| --- | --- |
| *Subjective:* <br> - Patient is a 34-year-old male with a 10-year history of GERD. <br> - Reports dysphagia and odynophagia localized to the bottom of the throat and upper chest, rated 6/10 in intensity. <br> - Symptoms have progressively worsened over time. <br> - Dysphagia occurs with solid foods; patient needs to drink water after bites to swallow. <br> - Frequent burping and need to clear throat often. <br> - Occasionally feels like something is stuck in throat when not eating. <br> - No issues with liquids, soft foods alleviate symptoms. <br> - Lessened appetite but no weight loss. <br> - Moderate alcohol use. <br> - No significant past illnesses, major injuries, surgeries, psychiatric problems, hospitalizations, or drug allergies. <br> - Currently taking a PPI with no recent changes. <br> - Family history: Mother has cancer, grandfather had heart disease. <br> - Works with a computer, no unusual exposures or stresses. <br> - No cardiovascular, respiratory, urinary, or neurological issues. <br><br> *Objective:* <br> - (Pending physical examination findings and test results) <br><br> *Assessment:* <br> 1. Gastroesophageal Reflux Disease (GERD) with possible Esophagitis or Stricture <br> 2. Esophageal Cancer <br> 3. Eosinophilic Esophagitis <br> 4. Achalasia <br><br> *Plan:* <br> 1. Laboratory Tests: <br> - Complete blood count (CBC) <br> - Serum eosinophils <br> - Liver function tests <br> 2. Imaging Studies: <br> - Barium Swallow X-ray <br> - Upper Endoscopy (EGD) with biopsy <br> - CT scan of the chest and abdomen <br> 3. Medications: <br> - Continue current PPI, consider increasing dose (e.g., Omeprazole 40 mg once daily) <br> - Topical corticosteroids if eosinophilic esophagitis is confirmed (e.g., Fluticasone 220 mcg swallowed twice daily for 8 weeks) <br> 4. Operations or Procedures: <br> - Endoscopic dilation if stricture is found <br> - Biopsy of any suspicious lesions during endoscopy | ID Verification: <br> Name, Date of Birth, present location, and home address verified, and I am licensed in the state the patient is currently located. <br><br> HPI: <br> See intake <br> Additional details not noted in intake: Food getting stuck in the upper esophagus has been going on for about one month. He has been take omprazole (sic) PRN <br><br> PMH: <br> See intake <br> Additional details not noted in intake: Long history of GERD <br><br> PSH: <br> Medications: Medications reviewed with patient. <br> Patient reports taking the following medications: Omeprazole and Pepcid <br> Allergies:No known drug allergies <br> Patient preferred pharmacy: <br><br> EXAM <br> The patient was evaluated via Video. <br> GEN: Alert and oriented x3; No acute distress; Appropriate mood and affect; <br> RESP: Speaking in full sentences without pause;Not coughing; No gasping or obvious signs of respiratory distress; <br> PSYCH: Good judgement and insight; <br><br> Assessment: <br> GERD with esophageal stricture <br><br> Plan: <br> Additional plan discussed with patient: Needs EGD, take the omeprazole on a daily bases <br> I recommend that the patient be seen: in the next 24-48 hours with their primary care provider. |

Case 2: This SOAP note pair received the following scores: top diagnosis: concordant; top 4 diagnoses: concordant; treatment plan alignment: concordant; CSS 8/10.

| Doctronic SOAP Note | Human Clinician SOAP note |
|---|---|
| **SOAP Note**<br><br>*Subjective*<br>- Patient is a 66-year-old woman.<br>- Chief complaint: Sore throat and thick white coating on throat for 1 week.<br>- Soreness intensity: 4/10.<br>- No associated symptoms such as fever, difficulty swallowing, swollen lymph nodes, changes in voice, hoarseness, unusual fatigue, weight changes, or skin rashes.<br>- Self-treatment with peroxide provided no relief.<br>- No history of similar symptoms in the past.<br>- Relevant medical history: High blood pressure.<br>- Current medications: Valsartan, HCTZ, Levothyroxine.<br>- No recent changes in medications or doses.<br>- No known allergies.<br>- Up to date with vaccinations.<br>- No use of tobacco, alcohol, or recreational drugs.<br>- No family history of major illnesses.<br>- Occupation: Nurse.<br><br>*Objective*<br>- No physical examination findings available since this is self-reported information.<br><br>*Assessment*<br>1. Oral Candidiasis (Thrush)<br>2. Strep Throat<br>3. Leukoplakia<br>4. Lichen Planus<br><br>*Plan*<br>1. **Laboratory Tests:**<br>   - Throat Swab and Culture to identify fungal organisms like Candida or rule out bacterial causes like Group A Streptococcus.<br>   - Serum Glucose Levels to rule out diabetes, which can predispose to fungal infections.<br>2. **Medications:**<br>   - For Oral Candidiasis (if confirmed):<br>   - Nystatin Oral Suspension: 100,000 units/mL, 4-6 mL swished and swallowed four times daily for 7-14 days.<br>   - For Strep Throat (if culture positive):<br>   - Penicillin V: 500 mg orally twice daily for 10 days.<br>   - For Symptomatic Relief:<br>   - Ibuprofen: 400 mg every 6 hours as needed for throat pain.<br>3. **Operations or Procedures:**<br>   - None immediately indicated based on current symptoms and differential diagnosis. | ID Verification:<br>Name, Date of Birth, present location, and home address verified, and I am licensed in the state the patient is currently located.<br><br>HPI:<br>See intake<br>Additional details not noted in intake: Recently was taking Flagyl last week for BV.<br>White patches on the tongue, cheeks, gums, or palate. States she can scrape it off.<br>Denies fever, chills, swollen lymph nodes or other signs of strep throat.<br><br>PMH:<br>See intake<br><br>PSH:<br>See intake<br><br>Medications:<br>Medications reviewed with patient.<br><br>Allergies:<br>No known drug allergies<br><br>Patient preferred pharmacy: Walgreens<br><br>EXAM:<br>The patient was evaluated via Video.<br><br>GEN: Alert and oriented x3; No acute distress; Appropriate mood and affect;<br><br>Assessment:<br>Oral Thrush<br><br>Plan: Nystatin Oral Suspension: 100,000 units/mL, 4-6 mL swished and swallowed four times daily for 14 days.<br>F/u with PCP if no improvement. |

Case 3: This SOAP note pair received the following scores: top 1: not concordant, top 4: concordant, treatment plan alignment: not concordant, CSS 3/10, The AI note was deemed superior by the human expert.

| Doctronic SOAP Note | Human Clinician SOAP note |
|---|---|
| *Subjective*:<br>- 42-year-old woman presents with sharp and constant left flank pain.<br>- Pain started recently and has become more concentrated.<br>- Pain intensity rated 7 out of 10.<br>- Laying down makes the pain worse.<br>- Associated symptom: frequent urination.<br>- No changes in urine appearance, smell, or amount.<br>- Recently completed a course of Bactrim for a UTI (7 days ago).<br>- No history of similar symptoms or past urological issues.<br>- No known drug allergies.<br>- No use of tobacco, alcohol, or recreational drugs.<br>- No significant family history of major illnesses.<br>- Not currently working.<br>- Up to date on preventive health measures.<br>- No recent weight loss, night sweats, fatigue, or changes in appetite.<br><br>*Objective*:<br>- No physical examination findings provided.<br>- No lab results provided.<br>- No imaging studies provided.<br><br>*Assessment*:<br>1. Renal Calculi (Kidney Stones)<br>2. Pyelonephritis<br>3. Interstitial Cystitis<br>4. Gynecological Issues (e.g., Ovarian Cyst)<br><br>*Plan*:<br>1. **Laboratory Tests**:<br>- Urinalysis and urine culture to check for infection, blood, or crystals.<br>- Complete Blood Count (CBC) to evaluate for systemic infection.<br>- Serum Creatinine and Blood Urea Nitrogen (BUN) to assess kidney function.<br>2. **Imaging Studies**:<br>- Non-contrast CT Scan of the abdomen and pelvis to detect kidney stones and other potential causes of flank pain.<br>- Renal Ultrasound to visualize the kidneys and urinary tract.<br>3. **Medications**:<br>- Ibuprofen, 400 mg every 6 hours as needed for pain.<br>- If pyelonephritis is confirmed: Ciprofloxacin, 500 mg twice daily for 7-14 days.<br>4. **Procedures**:<br>- Ureteroscopy or Extracorporeal Shock Wave Lithotripsy (ESWL) if kidney stones are detected and causing significant pain or obstruction. | Verification:<br>Name, Date of Birth, present location, and home address verified, and I am licensed in the state the patient is currently located<br><br>Subjective:<br>HPI:<br>See intake<br>Chief Complaint: flank pain<br>Onset: few days<br>Region/Radiation: just in low L back<br>Severity: 7/10<br>Patient denies blood, fever, other urine signs and symptoms save a little frequency<br>did just finish Bactrim DS for a UTI<br><br>Pertinent past Medical/Surgical History: See intake<br>Current Medications/Supplements/Drugs: See intake<br>Medication Allergies: NKDA<br><br>Objective:<br>Visit modality:<br>Video Visit<br><br>EXAM<br>The patient was evaluated via Video.<br>GEN: Alert and oriented x3; No acute distress; Appropriate mood and affect;<br>RESP: Speaking in full sentences without pause;Not coughing; No gasping or obvious signs of respiratory distress;<br><br>PSYCH: Good judgement and insight;<br><br>Assessment:<br>flank pain<br><br>Plan: watch for now and use Ibuprofen - advised on watching for blood, fever, return of UTI signs and symptoms |

Case 4: This SOAP note pair received the following scores: Top 1: Not-concordant, Top 4: Concordant, Treatment Plan Alignment: Concordant, CSS 8/10, Neither note was deemed superior by the expert reviewer

| Doctronic SOAP Note | Human Clinician SOAP note |
|---|---|
| *Subjective*<br>- 41-year-old female.<br>- Chief complaints: sore throat, upset stomach, and headache.<br>- Sore throat: Intensity 5/10, began 2 days ago, radiates to ears, worse in the morning, associated with difficulty swallowing in the morning and chills.<br>- Upset stomach: Uneasy feeling with loose stools and stomach cramping, intensity 4/10, began 2 days ago.<br>- Headache: Generalized across forehead and head, intensity 2/10, began 2 days ago.<br>- Additional symptom: Fatigue.<br>- No past medical history of major illnesses, surgeries, or hospitalizations.<br>- No current medications or recent changes to medications.<br>- No known allergies to medications, foods, or other substances.<br>- Up-to-date with vaccinations and regular health check-ups.<br>- No use of alcohol, tobacco, or recreational drugs.<br>- No family history of major illnesses.<br>- Occupation: Realtor.<br><br>*Objective*<br>- Patient appears in general good health but reports feeling unwell.<br>- No physical examination data provided (to be completed by healthcare provider).<br><br>*Assessment*<br>1. Viral Pharyngitis (Common Cold)<br>2. Gastroenteritis<br>3. Streptococcal Pharyngitis (Strep Throat)<br>4. Mononucleosis (Infectious Mononucleosis)<br><br>*Plan*<br>1. **Laboratory Tests:**<br>   - Rapid Strep Test and Throat Culture to check for group A Streptococcus infection.<br>  - Monospot Test for infectious mononucleosis.<br>2. **Medications:**<br>  - Acetaminophen (Tylenol), 500 mg every 6 hours as needed for pain to alleviate headache and sore throat discomfort.<br>  - Amoxicillin, 500 mg three times a day for 10 days if strep throat is confirmed.<br>  - Oral Rehydration Solutions to manage dehydration from loose stools and upset stomach. | HPI:<br>See intake<br>Additional details not noted in intake:<br><br>Started symptoms on Sunday with fatigue, headache. Daughter was diagnosed with strep on Friday. Yesterday/today has sore throat that is painful to swallow. Symptoms suddenly began. Denies fever. Tender/swollen tonsils/lymph nodes. Redness to back of throat. Denies cough. Negative Covid test.<br><br>Centor score 3 with positive exposure - will treat for strep.<br><br>Medications:<br>Medications reviewed with patient.<br><br>Allergies:<br>No known drug allergies<br><br>Patient preferred pharmacy: Walgreens<br><br>EXAM<br>The patient was evaluated via Video.<br><br>GEN:<br><br>RESP: Speaking in full sentences without pause;<br><br>Assessment:<br>Strep Pharyngitis<br><br>Plan:<br>Reviewed plan of care from intake with patient. Follow up instructions given. Pt verbalized understanding and agreed with plan |

# Appendix 2: Verbatim Prompts

Prompt 1: Consistency of Top 4 Diagnoses

## Your Role

You are an LLM that functions as a clinical evaluation assistant tasked with comparing diagnoses located within two SOAP notes. I want you to determine if any of the diagnoses from one SOAP note match or are clinically consistent with any of the diagnoses from the other note. For example, if one note has 4 diagnosis, and the other note has 1 diagnosis, determine if any of the 4 diagnoses from one SOAP note match or are consistent with the 1 diagnosis from the other SOAP note.

To complete the task, follow these steps.

1. Read both SOAP notes
2. Identify the listed diagnosis in each SOAP note. For this task, only use explicitly stated diagnosis. Diagnoses are only found in the assessment section of the SOAP note.
3. Compare all of the diagnosis from one SOAP note to all of the diagnosis in the other SOAP note.
4. Determine if 1 or more diagnosis from one note matches or is clinically consistent with 1 or more diagnosis from the other note.

## Evaluation Criteria

The diagnoses are considered to be "clinically consistent" if:

1. The primary diagnosis from one note is the same as or clinically similar to at least one of the diagnoses (primary or secondary) in the other note.
2. A diagnosis from one SOAP note is a more specific subtype of any diagnosis in the other SOAP note. For example, a diagnosis of "Lower Back Pain" in one SOAP note would be clinically consistent with a diagnosis of "Sciatica" in another SOAP note.
3. The diagnoses, while using different terminology, refer to the same underlying clinical condition. For example, a diagnosis of "Eczema" in one SOAP note would be clinically consistent with a diagnosis of "Atopic Dermatitis" in another SOAP note.
4. The diagnoses are overlapping conditions in the same body system with different wording (e.g., "Sinusitis" vs. "Allergic Rhinitis with Sinusitis")
5. One diagnosis is directly related to or a complication of another diagnosis. (e.g., "gallstones" vs. "cholecystitis" or "leg swelling" vs. "deep vein thrombosis").

## Examples of Inconsistent Diagnoses

- Completely different body systems without clear connection (e.g., "migraine" vs. "plantar fasciitis")
- Contradictory diagnoses for the same symptoms (e.g., "viral pneumonia" vs. "congestive heart failure" for the same presentation)
- Diagnoses that would lead to fundamentally different treatment approaches

## Output Format

- If the SOAP notes are clinically consistent respond with this exact phrase: <001>
- If the SOAP notes are NOT clinically consistent respond with this exact phrase: <000>
- You must respond with only the code.

Remember that your goal is to evaluate if the notes would lead to similar clinical treatment approaches, not whether they are identical in every detail.

Prompt 2: Consistency of the Top Diagnosis

**Your Role**

You are an LLM that functions as a clinical evaluation assistant tasked with comparing diagnoses located within two SOAP notes. I want you to determine if the primary diagnosis from one SOAP note matches or is clinically consistent to the primary diagnoses from the other note.

To complete the task, follow these steps.

1. Read both SOAP notes.
2. Identify the listed diagnosis in each SOAP note. For this task, only use explicitly stated diagnosis. Diagnoses are only found in the assessment section of the SOAP note.
3. Compare all of the diagnosis from one SOAP note to all of the diagnosis in the other SOAP note.
4. Determine if the primary or top diagnosis from one note matches or is clinically consistent with the primary diagnosis from the other note.

The diagnoses are considered to be "clinically consistent" if:

1. The primary diagnosis from one note is the same as or clinically similar to the primary diagnoses in the other note
2. The primary diagnosis from one SOAP note is a more specific subtype of the primary diagnosis in the other SOAP note. For example, a diagnosis of "Lower Back Pain" in one SOAP note would be clinically consistent with a diagnosis of "Sciatica" in another SOAP note.
3. The primary diagnoses, while using different terminology, refer to the same underlying clinical condition. For example, a diagnosis of "Eczema" in one SOAP note would be clinically consistent with a diagnosis of "Atopic Dermatitis" in another SOAP note.
4. The primary diagnoses are overlapping conditions in the same body system with different wording (e.g., "Sinusitis" vs. "Allergic Rhinitis with Sinusitis")
5. One diagnosis is directly related to or a complication of another diagnosis. (e.g., "gallstones" vs. "cholecystitis" or "leg swelling" vs. "DVT")

**Examples of Inconsistent Diagnoses**

- Completely different body systems without clear connection (e.g., "Migraine" vs. "Plantar Fasciitis")
- Contradictory diagnoses for the same symptoms (e.g., "Viral Pneumonia" vs. "Congestive Heart Failure" for the same presentation)
- Diagnoses that would lead to fundamentally different treatment approaches

**Output Format**

- If the SOAP notes are clinically consistent respond with this exact phrase <001>
- If the SOAP notes are NOT clinically consistent respond with this exact phrase <000>
- You must respond with only the code.

Remember that your goal is to evaluate if the notes would lead to similar clinical treatment approaches, not whether they are identical in every detail.

Prompt 3: Consistency of the Treatment Plan

**Your Role**

You are a clinical evaluation assistant tasked with comparing treatment plans between two SOAP notes to determine if they are clinically consistent with each other. Your job is to carefully analyze both treatment approaches and determine whether they represent compatible clinical management strategies that would lead to similar therapeutic outcomes. For the purposes of this comparison the treatment plan is usually contained in the final section of the SOAP note. Treatment plans can include, confirmation tests and imaging studies, medications, procedures, home care or ancillary services such as physical therapy.

**Evaluation Criteria**

Two SOAP notes are considered to have **clinically consistent treatment plans** if:
1. The core therapeutic approach is similar (e.g., both recommend physical therapy, medication management, or surgical intervention).
2. One treatment plan recommends a test or imaging study to confirm the diagnosis prior to treatment and the other does not, but they both agree on the final treatment approach.
3. The specific interventions, while possibly different in details, address the same underlying clinical needs.
4. The treatment modalities are recognized alternatives for the same condition.
5. One plan includes all key elements of the other plan plus additional elements (more comprehensive but includes the same core approach).
6. One note has very limited treatment plans such as a phrase or one sentence and the other has a longer treatment plan, but they both provide similar treatment approaches.

**Examples of Consistent Treatment Plans**
- Similar medication classes with different specific drugs (e.g., "Ibuprofen 600 mg three times a day" vs. "Naproxen 500 mg twice a day"; both are nonsteroidal anti-inflammatory drugs (NSAIDS)).
- Equivalent physical interventions (e.g., "Physical therapy focusing on lumbar strengthening" vs. "Home exercise program for core strengthening").
- Stepped care approaches covering the same therapies (e.g., "Try heat therapy, then NSAID, then consider PT referral" vs. "PT 2x weekly with home heat application and as-needed NSAID").
- Similar monitoring approaches (e.g., "Follow-up in 4 weeks with blood pressure log" vs. "Monitor blood pressure daily and return in 1 month for reassessment")

**Examples of Inconsistent Treatment Plans**
- Plans requiring fundamentally different approaches (e.g., "conservative management with NSAIDs" vs. "immediate surgical intervention").
- Contradictory interventions (e.g., "strict bed rest for 72 hours" vs. "maintain normal activity levels and avoid rest").
- Treatment plans addressing entirely different therapeutic goals.
- One plan recommending critical interventions that are explicitly avoided in the other plan.

**Analysis Process**
1. Extract all treatment elements from both SOAP note.
2. Categorize treatments by type (e.g., medications, physical therapies, procedures, lifestyle modifications, follow-up).
3. Compare similar categories across both notes.
4. Determine whether the core therapeutic approach is preserved between notes.
5. Determine whether any critical contradictions exist between the plans.

6. Provide your reasoning, citing specific therapeutic relationships.

**Output Format**
- Begin with a side-by-side comparison of all diagnoses from both notes.
- If the SOAP notes treatment plans are clinically consistent, respond with this exact phrase: <001>.
- If the SOAP notes treatment plans are NOT clinically consistent, respond with this exact phrase: <000>.
- You must respond with only the code.

Remember that your goal is to evaluate whether the treatment plans would lead to similar therapeutic outcomes, not whether they are identical in every detail. Treatment plans may vary in specific medications, dosages, or techniques while still maintaining consistency in the overall therapeutic approach.

Prompt 4: Clinical Similarity Score (CSS)

**Your Role**

You are a clinical documentation reviewer. Your task is to compare two SOAP notes-one generated by Doctronic and the other by a human clinician-and evaluate them based on the following four criteria. Provide your output in following formats:

Similarity: X/10 | Complexity: Y/10 | Co-morbidity: Yes/No | ICD: [Main Clinical Condition]

Difference: One sentence summarizing the key clinical or documentation difference between the two notes

**Scoring Criteria**

1. Similarity: Rate how similarly a physician would understand the patient's condition from both SOAP notes.
   - 10: Nearly identical in clinical impression and documentation.
   - 7-9: Clinically very aligned with minor phrasing or detail differences.
   - 4-6: Moderate differences in emphasis, phrasing, or included symptoms.
   - 1-3: Major differences in clinical interpretation or diagnostic focus.
   - 0: Completely different clinical portrayals.
2. Complexity: Rate the medical complexity of the case.
   - 0-2: Very common and simple conditions (e.g., seasonal flu, tension headache).
   - 3-5: Mild-to-moderate complexity (e.g., hypertension with lifestyle considerations).
   - 6-8: Significant complexity or multi-system involvement (e.g., type 1 diabetes, chronic obstructive pulmonary disease).
   - 9-10: Rare, severe, or highly specialized conditions (e.g., polymyositis, amyotrophic lateral sclerosis).
3. Co-morbidity
   - Indicate whether the patient has one or more additional chronic or acute conditions that are clinically relevant to the case.
   - ICD condition: Identify the main diagnosis using ICD-compatible terminology (e.g., "Acute Sinusitis", not "sinus infection" or ICD-10 code).

**Output Format Example**

Similarity: 8/10 | Complexity: 3/10 | Co-morbidity: No | ICD: acute viral rhinitis.

Difference: Doctronic's note emphasized upper respiratory symptoms, while the human clinician highlighted general fatigue without mention of nasal congestion.

**Instructions**

Read both SOAP notes thoroughly. Focus on capturing the core clinical impression of each note and then assess the extent to which the diagnostic and management reasoning in the two notes align. Use the similarity score to quantify alignment, the complexity score to characterize clinical difficulty, and co-morbidity to flag additional medical conditions. Summarize the main diagnosis using standard ICD-compatible language, and finally, articulate the key clinical or stylistic difference in one sentence.

# Appendix 3: ICD-10 Codes and Frequency

This is a list of all the ICD-10 Codes that appeared more than once. ICD-10 codes represent the primary diagnosis taken from the AI generated SOAP note with the frequency (F) of that code over 500 visits.

| ICD-10 | F | ICD-10 | F | ICD-10 | F |
|---|---|---|---|---|---|
| Acute Sinusitis | 25 | Acute Upper Respiratory Infection | 3 | Irritable Bowel Syndrome (IBS) | 2 |
| Influenza | 22 | Asthma with (acute) exacerbation | 3 | Major Depressive Disorder | 2 |
| Acute Bronchitis | 18 | Atopic Dermatitis | 3 | Multiple Sclerosis | 2 |
| Acute Bacterial Sinusitis | 16 | Attention-Deficit/Hyperactivity Disorder | 3 | Major Depressive Disorder | 2 |
| Viral Upper Respiratory Infection | 13 | Chronic Fatigue Syndrome | 3 | Multiple Sclerosis | 2 |
| Generalized Anxiety Disorder | 10 | Dermatitis | 3 | Onychomycosis | 2 |
| Acute Pharyngitis | 9 | Essential (Primary) Hypertension | 3 | Paronychia | 2 |
| Acute Viral Upper Respiratory Infection | 9 | Gastroenteritis | 3 | Premature Ejaculation | 2 |
| Bacterial Vaginosis | 9 | Low Back Pain | 3 | Pruritus Ani | 2 |
| Viral Gastroenteritis | 9 | Urethritis | 3 | Rash and other nonspecific skin eruption | 2 |
| Dental Abscess | 8 | Acute Cystitis | 2 | Rheumatoid Arthritis | 2 |
| Cellulitis | 7 | Acute Gastroenteritis | 2 | Sinusitis | 2 |
| Streptococcal Pharyngitis | 7 | Acute Viral Pharyngitis | 2 | Systemic Lupus Erythematosus | 2 |
| Essential Hypertension | 6 | Chest Pain, Unspecified | 2 | Tinea Cruris | 2 |
| Gastroesophageal Reflux Disease | 6 | Chronic Lumbar Radiculopathy | 2 | Viral Sinusitis | 2 |
| Oral Candidiasis | 6 | Chronic Obstructive Pulmonary Disease | 2 | Vulvovaginal Candidiasis | 2 |
| Viral Pharyngitis | 6 | Chronic Prostatitis/Chronic Pelvic Pain Syndrome | 2 | | |
| Bacterial Conjunctivitis | 5 | Contact Dermatitis | 2 | | |
| Chronic Sinusitis | 5 | Costochondritis | 2 | | |
| Hypothyroidism | 5 | COVID-19 | 2 | | |
| Urinary Tract Infection | 6 | Eustachian Tube Dysfunction | 2 | | |
| Acute Viral Sinusitis | 4 | Genital Herpes (Herpes Simplex Virus) | 2 | | |
| Balanitis | 4 | Gout | 2 | | |
| Cervical Radiculopathy | 4 | Herpes Zoster | 2 | | |
| Erectile Dysfunction | 4 | Hordeolum (Stye) | 2 | | |
| Urinary Tract Infection (UTI) | 4 | Hypertensive Crisis | 2 | | |